\documentclass[aps,showpacs,amsmath,amssymb,twocolumn,pra,superscriptaddress,notitlepage]{revtex4-1}

\usepackage[dvipdfmx]{graphicx}
\usepackage{latexsym}
\usepackage{amsmath,amssymb,amsthm,mathrsfs,amsfonts,dsfont}
\usepackage{subcaption}
\usepackage{braket}
\usepackage{bm}
\usepackage{enumerate}
\usepackage{color}
\usepackage{algorithm}
\usepackage{algorithmic}
\usepackage{comment}
\usepackage{appendix}
\usepackage{here}
\usepackage{qcircuit}
\usepackage{tikz}
\usepackage{siunitx} 
\usepackage[colorlinks,linkcolor=blue,citecolor=blue]{hyperref}

\newcommand{\black}[1]{\textcolor{black}{#1}}

\def\Underline{\setbox0\hbox\bgroup\let\\\endUnderline}
\def\endUnderline{\vphantom{y}\egroup\smash{\underline{\box0}}\\}

\usepackage{ulem}

\definecolor{color1}{rgb}{0,0.5,0}

\definecolor{color2}{rgb}{0.5,0.5,0}

\definecolor{color3}{rgb}{0,0.5,0.5}

\begin{document}
\title{Leveraging hardware-control imperfections for error mitigation via generalized quantum subspace}

\author{Yasuhiro Ohkura}
\email{yasuhiro.ohkura.1997@gmail.com}
\affiliation{Graduate School of Media and Governance, Keio University SFC, Fujisawa, Kanagawa 252-0882 Japan}

\author{Suguru Endo}
\affiliation{NTT Computer and Data Science Laboratories, NTT Corporation, Musashino 180-8585, Japan}
\author{Takahiko Satoh}
\affiliation{Keio University Quantum Computing Center, Yokohama, Kanagawa 223-8522 Japan}
\affiliation{Graduate School of Science and Technology, Keio University, Yokohama, Kanagawa 223-8522 Japan}
\author{Rodney Van Meter}
\affiliation{Faculty of Environment and Information Studies, Keio University SFC, Fujisawa, Kanagawa 252-0882 Japan}
\author{Nobuyuki Yoshioka}
\email{nyoshioka@ap.t.u-tokyo.ac.jp}
\affiliation{Department of Applied Physics, University of Tokyo, 7-3-1 Hongo, Bunkyo-ku, Tokyo 113-8656, Japan}
\affiliation{Theoretical Quantum Physics Laboratory, RIKEN Cluster for Pioneering Research (CPR), Wako-shi, Saitama 351-0198, Japan}
\affiliation{JST, PRESTO, 4-1-8 Honcho, Kawaguchi, Saitama, 332-0012, Japan}

\thanks{Y.O. and N.Y. contributed to this work equally.}

\begin{abstract} 
In the era of quantum computing without full fault-tolerance, it is essential to suppress noise effects via the quantum error mitigation techniques to enhance the computational power of the quantum devices. 
One of the most effective noise-agnostic error mitigation schemes is the generalized quantum subspace expansion (GSE) method, which unifies various mitigation algorithms under the framework of the quantum subspace expansion. 
Specifically, the fault-subspace method, a subclass of GSE method, constructs an error-mitigated quantum state with copies of quantum states with different noise levels. 
However, from the experimental aspect, it is nontrivial to determine how to reliably amplify the noise so that the error in the simulation result is efficiently suppressed. 
In this work, we explore the potential of the fault-subspace method by leveraging the hardware-oriented noise: intentional amplification of the decoherence, noise boost by insertion of identity, making use of crosstalk, and probabilistic implementation of noise channel.
We show the validity of our proposals via both numerical simulations with the noise parameters reflecting those in quantum devices available via IBM Quantum, and also \textcolor{black}{demonstration}s performed on the hardware.
\end{abstract}

\maketitle

\section{Introduction}
To perform practical quantum computing, it is vital to suppress the computational errors that root from the decoherence of quantum registers due to coupling with the external environments, fabrication imperfection, and erroneous operation instructions. 
A fundamental solution is the realization of fault-tolerant quantum computers that executes computation accompanied with the  quantum error correction~\cite{shor1995scheme, knill1998resilient, nielsen2002quantum}. However, the quantum error correction often requires a large number of physical qubits to encode quantum information into logical qubits; for example, it has been estimated that  hundreds of logical qubits and millions to billions of logical gates are required to achieve advantages over classical algorithms~\cite{Babbush2018, Gidney2021howtofactorbit, lee_prxq_2021, goings_pnas_2022, yoshioka2022hunting, beverland2022assessing}. 
Despite the recent landmark achievements of break-even in experiments for small QEC codes~\cite{sivak2022real}, current quantum devices are still far from such highly complex quantum architectures~\cite{preskill2018quantum}. Even as the quantum error correction experiments have become more mature, if the logical rate does not reach the required level by the algorithm, we still need a methodology to reduce detrimental impacts of noise~\cite{piveteau2021error, lostaglio2021error, suzuki2022quantum, tsubouchi2023virtual}. Quantum error mitigation (QEM) is a body of methods for such a practical and indispensable objective.

QEM refers to noise-reduction techniques that rely on the classical postprocessing of measurement outcomes~\cite{temme2017error, li2017efficient, endo2021hybrid}. 
One of the earliest proposals that has been utilized heavily in quantum devices is the probabilistic error cancellation~\cite{temme2017error, endo2018practical, berg2022probabilistic}, which constructs the inverse operations of noise with additional single-qubit operations and classical-post processing of measurement outcomes. 
While the QEM techniques do not replace the quantum error correction due to the growth of sampling overhead~\cite{takagi2022fundamental, tsubouchi2022universal, takagi2022universal, quek2022exponentially}, it has been recognized as an essential tool to make full use of non-fault-tolerant algorithms such as the variational quantum eigensolver~\cite{peruzzo2014variational, kandala2019error, yoshioka_2020_dvqe}.

While the probabilistic error mitigation method heavily depends on the characterization of the circuit errors and is hence categorized as {\it error-aware} QEM methods, there is also a class of {\it error-agnostic} QEM methods that do not rely on such knowledge: zero-noise extrapolation~\cite{temme2017error}, (generalized) quantum subspace expansion~\cite{mcclean_2017, Yoshioka2021, yoshioka2022variational}, and virtual distillation (VD)~\cite{koczor2020exponential, huggins2020virtual}.
The VD method is designed to purify mixed states (hence also called purification-based QEM) using multiple copies of noisy quantum states so that the effect of the subdominant eigenvectors of the density matrix is suppressed drastically with respect to the number of copies.
This efficiently eliminates stochastic errors while it is completely susceptible to coherent errors.
The generalized quantum subspace expansion (GSE) method was proposed to overcome such an issue by constructing a variational quantum-classical hybrid ansatz that makes use of entangled measurements between multiple quantum states~\cite{Yoshioka2021}. 
As the name implies, GSE is also a strong generalization of quantum subspace expansion (QSE) in the sense that it incorporates the strength of the VD and QSE method allow efficient reduction of both stochastic and coherent errors. 

In GSE method, it is crucial to choose the appropriate construction of the subspace so that the variational ansatz reaches a better solution. One of the practical choices is to compose an expanded subspace from ``faults," namely a set of quantum states with different error rates.
The variational ansatz expanded with the ``fault subspace" has demonstrated that GSE method can efficiently leverage hardware imperfections, inheriting the advantage of the well-studied zero-noise extrapolation method. It is noteworthy that, unlike the case of zero-noise extrapolation, noise parameters do not need to be fully under control due to variational optimization. 
In Ref. \cite{Yoshioka2021}, the preparation of the fault subspace was done by boosting the error rate of each gate in an abstract manner. Therefore, it is important to consider the hardware-efficient preparation of the fault subspace. 

In this work, we explore new methods to induce additional noise to the noisy quantum states to generate expressive and hardware-friendly forms of fault subspace and study their effects on the quantum many-body simulations. In particular, we propose four flavors of fault subspace that leverage hardware imperfections with small experimental overhead.
The proposed methods are numerically demonstrated based on noise parameters reported for the quantum device \textit{ibm\_kawasaki}, which is one of the IBM Quantum Falcon r5.11 processors and one flavor that can be realized with minimal  effort is experimentally demonstrated on hardware provided therein.

The remainder of the present work is given as follows. 
In Sec.~\ref{sec:GSE_review}, we briefly review the techniques underlying the fault subspace of GSE method.
After introducing the hardware-oriented construction of the fault space in Sec.~\ref{sec:hardware_fault}, we present the numerical and experimental results in Sec.~\ref{sec:demonstration}.
Finally, we discuss the obtained results and future perspective of the current work in Sec.~\ref{sec:conclusion}.

\section{Formalism of Generalized Quantum Subspace Expansion Method}\label{sec:GSE_review}

GSE method is designed to mitigate errors in computing expectation values of physical observables by effectively simulating the following variational ansatz~\cite{Yoshioka2021}:
\begin{equation}\label{eqn:gse_ansatz}
    \rho_{\rm GSE} = \frac{P^\dag A P}{{\rm Tr}[P^\dag A P]},
\end{equation}
where $P = \sum_i c_i P_i~(c_i \in \mathbb{C})$ with $P_i$ being a generic operator, and $A$ is any positive-semidefinite Hermitian operator. 
Note that, once the coefficients $\{c_i\}_i$ are determined so that $\rho_{\rm GSE}$ well-approximates the target state, the quantum state
can be {\it effectively} simulated via classical-post processing of measurement outcomes.
Entangled measurements between multiple quantum states allow the coupler $P_i$ to take the following form:
\begin{equation}
    P_i = \sum_k \beta_k^{(i)} \prod_{l=1}^{L_k}W_{lk}^{(i)} \rho_{lk}^{(i)} V_{lk}^{(i)},
\end{equation}
where $\beta_{k}^{(i)} \in \mathbb{C}$, $\rho_{lk}^{(i)}$ is a quantum state, and $W_{lk}^{(i)}$ and $V_{lk}^{(i)}$ are considered to be unitary operators, while we may also allow non-unitary ones, e.g., via block-encoding, by use of ancillary qubits~\cite{berry_exponential_2014}. 
We assume that a measured physical observable $O=\sum_Q c_Q P_Q~(c_Q \in \mathbb{C}, P_Q\in \{I, X, Y, Z\}^{\otimes N})$ is represented by a linear combination of polynomially many Pauli operators. Then, the expectation value of the physical observable is expressed as 
\begin{equation}\label{eqn:observable}
    \braket{O} = \frac{\vec{c}^\dag \widetilde{O} \vec{c}}{\vec{c}^\dag S \vec{c}},
\end{equation}
where $\widetilde{O}_{ij} = {\rm Tr}[P_i^\dag A P_j O]$ and $\widetilde{S}_{ij} = {\rm Tr}[P_i^\dag A P_j]$. It is required that these quantities can be efficiently evaluated with quantum computers.

\black{When the target of the simulation is the ground state of a given Hamiltonian, in particular,}  $\vec{c}$ is optimized so that the energy expectation value is minimized. 
\black{This is performed under the Rayleigh-Ritz variational principle, which leads to} the following generalized eigenvalue problem:
\begin{equation}
    \widetilde{H}\vec{c} = E \widetilde{S} \vec{c},
\end{equation}
where $\widetilde{H}_{ij} = {\rm Tr}[P_i^\dag A P_j H]$ denotes the truncated Hamiltonian within the subspace spanned by the couplers, $\widetilde{S}_{ij} = {\rm Tr}[P_i^\dag A P_j]$ is the metric of the bases of the subspace, with $E$ denoting the approximated energy.

\begin{figure*}[ht]
\begin{center}
\resizebox{\hsize}{!}{\includegraphics{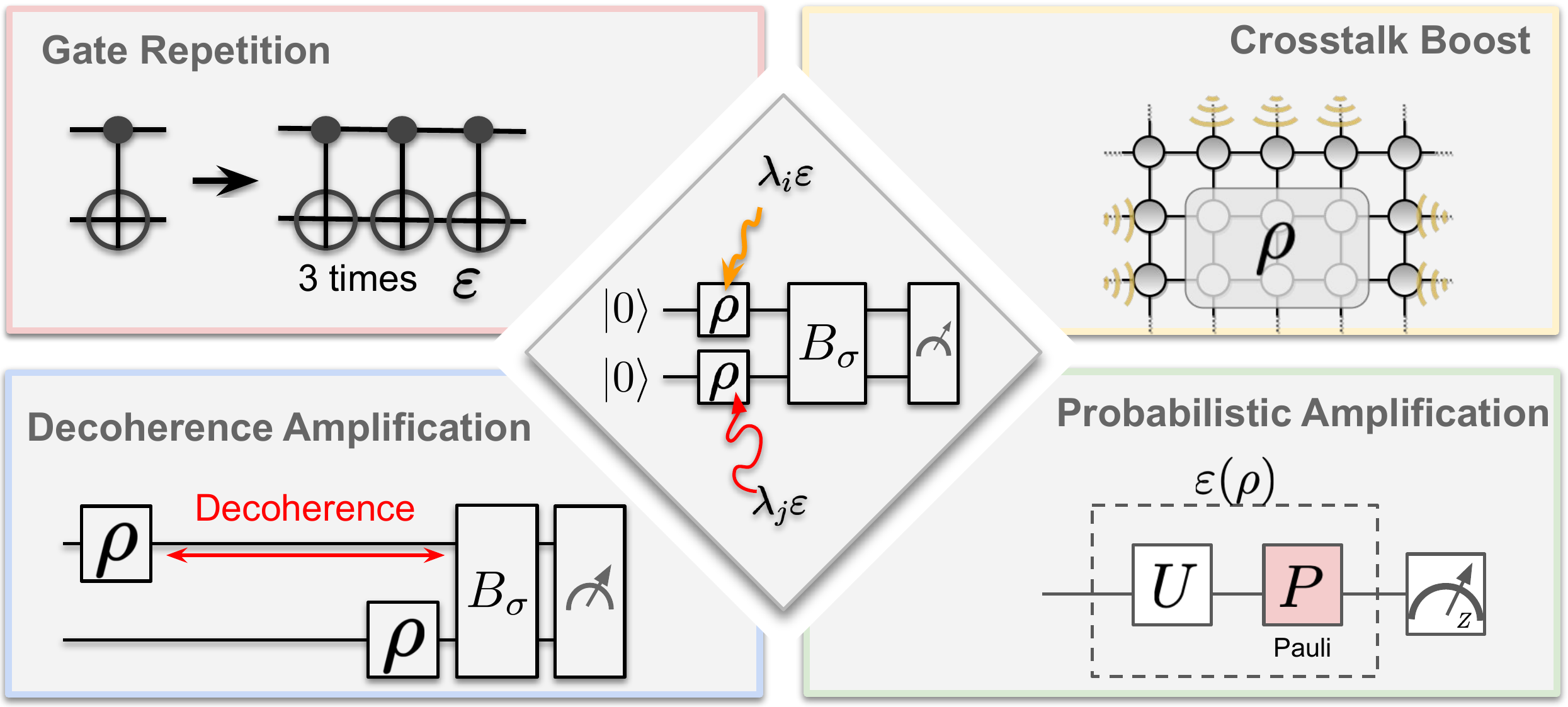}}
    \caption{
    Schematic picture of four practical implementations of GSE method that is designed to leverage imperfection of hardware instruction.
    \textbf{\textit{Decoherence Amplification: }}
    Just before the measurement operation, a certain amount of buffer time is inserted to induce thermal relaxation ($T_1$) and dephasing ($T_2$) in each qubit.
    \textbf{\textit{Gate Repetition: }}
    Amplifying the circuit noise via gate repetition of CX gate.
    In this example, two CX gates are canceled out, while the error is expected to increase by a factor of three.
    \textbf{\textit{Crosstalk Boost: }}
    By performing CX gates simultaneously on adjacent environments in the processor, we intentionally boost the error rate of CX gates via crosstalk.
    \textbf{\textit{Probabilistic Amplification: }}
    Artificially introduce depolarizing noise by stochastically inserting the Pauli gates right after every operation in the quantum circuit.
    }
\label{fig:leverraging_hardware_imperfection}
\end{center}
\end{figure*}

The fault-subspace method is a subclass of GSE method which employs error-modified quantum states as the couplers. 
To be concrete, we set
\begin{eqnarray}
    P_i &=& \rho(\varepsilon_i), \label{eq:fault_subspace_coupler}\\
    A&=& I/2^N, \label{eq:fault_subspace_A}
\end{eqnarray}
where $\{\varepsilon_i\}_i$ denotes a set of variable error rates that can be obtained by amplifying \black{(or possibly reducing via the probabilistic error cancellation)} the achievable error rate in the quantum circuit. 
The ansatz with Eqs.~\eqref{eq:fault_subspace_coupler},~\eqref{eq:fault_subspace_A} enables us to express the error mitigated state $\rho_{\rm GSE}\propto \bar{\rho}^\dag \bar{\rho}$ with $\bar{\rho}=\sum_i c_i \rho(\varepsilon_i)$, which involves
entangled measurements that for instance computes ${\rm Tr}[\rho_1 \rho_2] = {\rm Tr}[\Lambda (\rho_1 \otimes \rho_2)]$ using the derangement operator $\Lambda \ket{\phi_1}\ket{\phi_2} = \ket{\phi_2} \ket{\phi_1}$.

As discussed in Ref.~\cite{Yoshioka2021}, by choosing $\varepsilon_i$ to be governed by a homogeneous error stretch factor over the sequence of entire gate operations, the fault subspace inherit the advantage of both zero-noise extrapolation method and the virtual distillation method. 
Furthermore, it is crucial to point out that, although zero-noise extrapolation assumes the error-stretch factor to be completely controlled and hence sensitive to noise drift, i.e., an imperfection in noise characterization, such a deviation of the error model is absorbed into the coefficients $\{c_i\}_i$ which is a parameter optimized to encode the target state more accurately.

We emphasize that the well-controlled noise amplification is itself a highly nontrivial problem. 
In typical quantum devices, there are both temporal and spacial correlations in noise which are difficult to be captured by standard noise characterization techniques. 
This implies that the noise stretching may also involve amplification of uncharacterized errors, which leads to breakdown of the assumption in ordinary mitigation techniques (such as zero-noise extrapolation) that noise is well-described by time-independent Markovian noise.
Therefore, by exploring how to reliably boost noise based on hardware-level instructions, we expect to contribute not only to the community of QEM, but also to the quantum science technology that involves hardware characterization and quantum metrology as well. 

\section{Hardware-oriented fault subspace}\label{sec:hardware_fault}

In this section, we introduce four methods for highly practical experimental demonstration of GSE method based on the fault subspace: decoherence amplification, gate repetition, and crosstalk boost, and probabilistic amplification of errors. See Fig.~\ref{fig:leverraging_hardware_imperfection} for a graphical description.

\subsection{Decoherence Amplification}
One of the most common error source in quantum devices is the decoherence time of the quantum register. 
The decoherence time is commonly represented by two measures. One is $T_1$ which indicates the energy relaxation time, and the other is $T_2$ (or $T_2^*$) which is a metric of the duration of phase coherence.
In this hardware-oriented error boost method, we intentionally insert buffer time between operations to induce decoherence.
We may occasionally discriminate the method as GSE-DA method for the sake of convenience.

\black{Note that the concept of the decoherence amplification can be applied at any position of the quantum circuit since it does not require any additional hardware parameter tuning or calibration.
For the sake of simplicity, in the current work, we increase the buffer time between the last quantum gate and the measurement operation so that the circuit compilation is simplified.
Considering that the noise channels do not commute with quantum gates in general, one may construct various flavors of fault subspace via modification/optimization of the distribution of the buffer time along the entire circuit. 
Meanwhile, since the stability of GSE method is highly affected by the overlap between bases of subspace, we here prefer to avoid such a variation and focus on the single case with the highest implementability.
}

\subsection{Gate Repetition}
The gate repetition technique is a well-established technique for amplifying the effect of gate error by inserting a decomposed identity, given by a sequence of multiple gates as $\prod_n U_n = I$~\cite{dumitrescu2018cloud, tiron2020digital, setia2019reducing}. 
Since the insertion typically does not require any hardware calibration or circuit compiling, it is considered to be one of the simplest techniques to artificially boost the noise in a quantum circuit.

While we can in general choose any gate sets $\{U_n\}$ at any position as long as their product is identity, the common choice established in experiments is to make entangling gates (CX or CZ) redundant, which is expected to be the main error source in many experimental setups such as the superconducting qubits~\cite{Dumitrescu2018}. After choosing an entangling gate that satisfies $U^2 = I$, we insert an even number of entangling gates $U_1 = U_2 = \cdots U_{2K} = U$, so that the error from $U$ is approximately multiplied by a factor of $2K+1$. 

\subsection{Crosstalk Boost}
The crosstalk is one of the most detrimental errors in quantum processors~\cite{Krantz_2019,Ospelkaus2008}.
To be more concrete, crosstalk is the unwanted interaction between qubits that are not intended to couple with each other. The error typically arises when one of the connected (or nearby) qubits are interacting with other ones. When the fabrication/geometry of the quantum registers is not sufficiently isolating multi-qubit operations from each other, an entangling operation may simultaneously induce coupling between other qubit pairs.
Such a ``leak of coupling operation" is known to have a trade-off between the strength of qubit interaction and the magnitude of unwanted crosstalk noise~\cite{Gambetta2012,Sheldon2016}. 

The suppression of the crosstalk becomes a significant roadblock when we aim to develop larger quantum processors \cite{Sheldon2016,Rudinger2019,Harper2020}, and there are already several approaches designed for superconducting qubits.
In the case of quantum processors with tunable couplers including Google's Sycamore design~\cite{arute2019quantum}, we can tune qubit frequencies or control specific couplers so that the crosstalk is shut down \cite{Mundada2019,ding2020systematic}.
In fixed frequency qubit systems including superconducting qubit devices provided by IBM, on the other hand, the circuit scheduler must be designed so that concurrent execution of entangling operation between potentially cross-talking qubits is avoided as much as possible~\cite{Murali2020}.

\begin{figure}[ht]
    \begin{center}
    \resizebox{0.95\hsize}{!}{\includegraphics{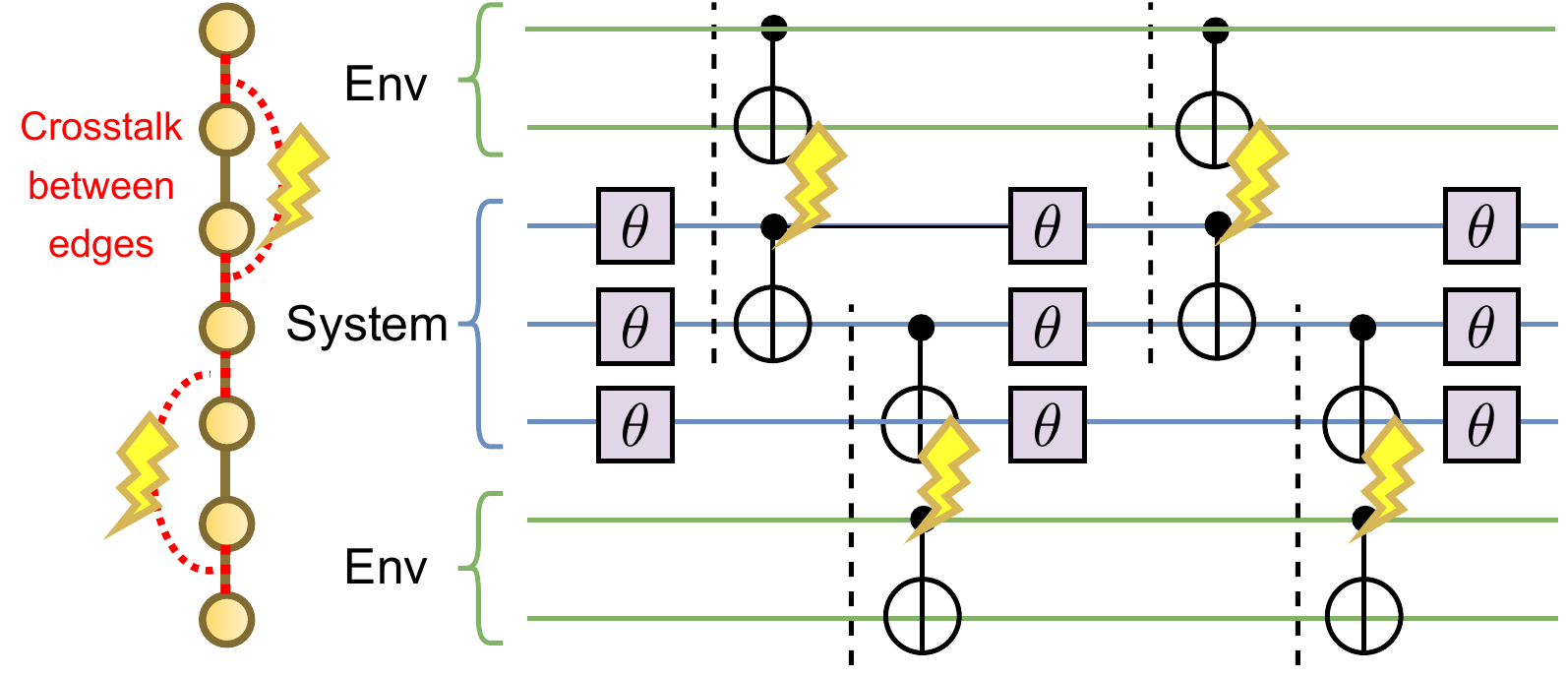}}
        \caption{
            A schematic illustration of the crosstalk boost method.
            The quantum circuit with three physical qubits as the simulation target system and four qubits of both ends are used as environments.
            By performing CX gates simultaneously on adjacent environments, we intentionally boost the error rate of CX gates via crosstalk.
        }
    \label{fig:crosstalkboost}
    \end{center}
\end{figure}

On the contrary, we here focus on intentionally {\it boosting} crosstalk errors by amplifying the interference between system qubits for the target quantum simulation and environment qubits solely for noise boost so that the fault subspace spanned by erroneous states is readily expanded. 
Figure~\ref{fig:crosstalkboost} shows an example that additionally generates crosstalk via cross-resonance gate operations. Here, we perform two-qubit gates on both the system qubits and the neighboring environmental qubits at the same time.
Note that the gate operation on the environment does not affect the quantum computation for the system qubit in the ideal setup, while in reality, we expect a boost in the ordinarily unwanted correlation with the external system.

\subsection{Probabilistic Amplification of Errors}
A relatively platform-independent way to intentionally boost errors is to implement the Kraus operator of noise channels in a probabilistic manner.
In other words, we may consider a set of relatively low-error unitaries $\{\mathcal{U}_k\}_k$ to generate a Kraus map
\begin{eqnarray}
    \mathcal{K}(\rho) = \sum_k p_k \mathcal{U}_k(\rho)\approx \sum_k p_k U_k\rho U_k^\dag,
\end{eqnarray}
where $\sum_k p_k = 1$ and $U_k U_k^\dag = I$ is the noiseless unitary operation. For instance, we may consider the entire set of Pauli operators $\{I, X, Y, Z\}^{\otimes N}$ to implement arbitrary Pauli errors.

Note that, when the noise channel itself is also a Pauli error with well-characterized amplitudes, such a probabilistic amplification may realize well-behaved control of the noise stretch factor.
This is in line with the original scenario considered in Ref.~\cite{Yoshioka2021}.
However, here we focus on leveraging native hardware errors and leave the current method as interesting future work.

\section{Demonstration of algorithm} \label{sec:demonstration}
\black{Now we proceed to provide the demonstration of the hardware-oriented GSE methods in both numerical and quantum device setups.
To clarify our capability of mitigating the error, we simulate the ground state of a given Hamiltonian from a noisy quantum circuit whose variational parameters are determined from classical optimization beforehand. To be concrete, we consider the one-dimensional transverse-field Ising (1d TFI) model with the Hamiltonian given as $H=-\sum_{r} Z_{r} Z_{r+1}+h \sum_{r} X_{r}$. Here, $X_r$ and $Z_r$ denote the $x$-and $z$-components of the Pauli matrix acting on the $r$-th site, and $h$ is the amplitude of the transverse magnetic field which is set to unity throughout this work.
We refer readers to Appendix.~\ref{app:ansatz} for the explicit form of the variational ansatz.
}

\subsection{Setup of numerical demonstration}
\label{subsec:numerics_setup}
Here we present the details regarding the numerical demonstration.
Our objective of running the numerical simulation here is to compare the hardware-oriented error mitigation techniques using noise models that capture the realistic features in actual hardware.
In this sense, it is instructive to describe how to concretely introduce noise for numerical demonstration in a hardware-oriented fashion.
The overall procedure can be summarized as follows:
\begin{enumerate}
    \item[Step] 1. Determine the high-level representation of the quantum circuit.
    \item[Step] 2. Transpile the circuit so that all the gates are expressed by elementary gates that are natively realized on the hardware.
    \item[Step] 3. Introduce noise channels for every native gate.
    \item[Step] 4. Simulate the noisy quantum state and the physical observables from the density matrix.
\end{enumerate}

Regarding Step 2, we assume that the native operations in the quantum circuits are given in accordance with IBM devices as $\sqrt{X}$, $R_z$, and CX gates:
\begin{eqnarray}
    \rm \sqrt{X} &=& \frac{1}{\sqrt{2}}\left(\begin{array}{cc}
e^{i \pi / 4} & e^{-i \pi / 4} \\
e^{-i \pi / 4} & e^{i \pi / 4}
\end{array}\right),\\
    R_z(\theta) &=& \left(\begin{array}{cc}
        1 & 0 \\
         0 &  e^{-i \theta}
    \end{array} \right), \\
    \mathrm{CX}&=&\left(\begin{array}{llll}
        1 & 0 & 0 & 0 \\
        0 & 1 & 0 & 0 \\
        0 & 0 & 0 & 1 \\
        0 & 0 & 1 & 0
        \end{array}\right).
\end{eqnarray}
Furthermore, we assume that the $R_z$ gates are operated only virtually by tuning the phase of the gate pulse; no actual operation is executed on the hardware. Such a framework is employed in platforms such as fixed-frequency transmon qubits.

Regarding Step 3, we assume that any $\sqrt{X}$ and CX gates are associated with the single-qubit depolarizing channel, and in addition, all qubits undergo decoherence caused by $T_1$ and $T_2$ effects. 
While the proposed methods are demonstrated under various noise levels, we fix the ratio between these quantities so that it reflects the actual noise characteristics of quantum hardware (See App.~\ref{app:noise}). 

Let us also make a remark on the errors accompanied with the entangled measurement.
 Concretely, the error in the basis transformation required for the entangled measurement (which involves swap/derangement operation along qubits) may totally deteriorate the result.
To see this, let us consider computing the following quantity via entangled measurement as
\begin{eqnarray}
    {\rm Tr}[\rho_1 \rho_2] = {\rm Tr}[\Lambda (\rho_1 \otimes \rho_2)],
\end{eqnarray}
where the derangement operator $\Lambda \ket{\phi_1}\ket{\phi_2}= \ket{\phi_2} \ket{\phi_1}$, or the swap operator for $M=2$ copy case, is diagonalized as follows,
\begin{eqnarray}
    \Lambda &=& \bigotimes_n \Lambda_{n, n+N} = \bigotimes_n (B_\sigma^\dag D B_\sigma),\\
    B_\sigma &=& \left(\begin{array}{cccc}
    1 & 0 & 0 & 0\\
    0 & 1/\sqrt{2} & - 1/\sqrt{2} & 0\\
    0 & 1/\sqrt{2} & 1/\sqrt{2} & 0\\
    0 & 0 & 0 & 1
    \end{array}\right), 
        D = {\rm diag}(1, -1, 1, 1).\nonumber
\end{eqnarray}
This indicates that we must operate unitary $B_\sigma$ on every qubit pair after all quantum gates in the circuit are executed.
In the numerical demonstration, we clarify the influence of errors present in the measurement-associated unitary circuit by comparing the results with and without noise in them.

\subsection{Numerical results} \label{subsec:numerics}
\begin{figure*}[t]
    \centering
        \begin{subfigure}{.32\hsize}
            \resizebox{\hsize}{!}{\includegraphics{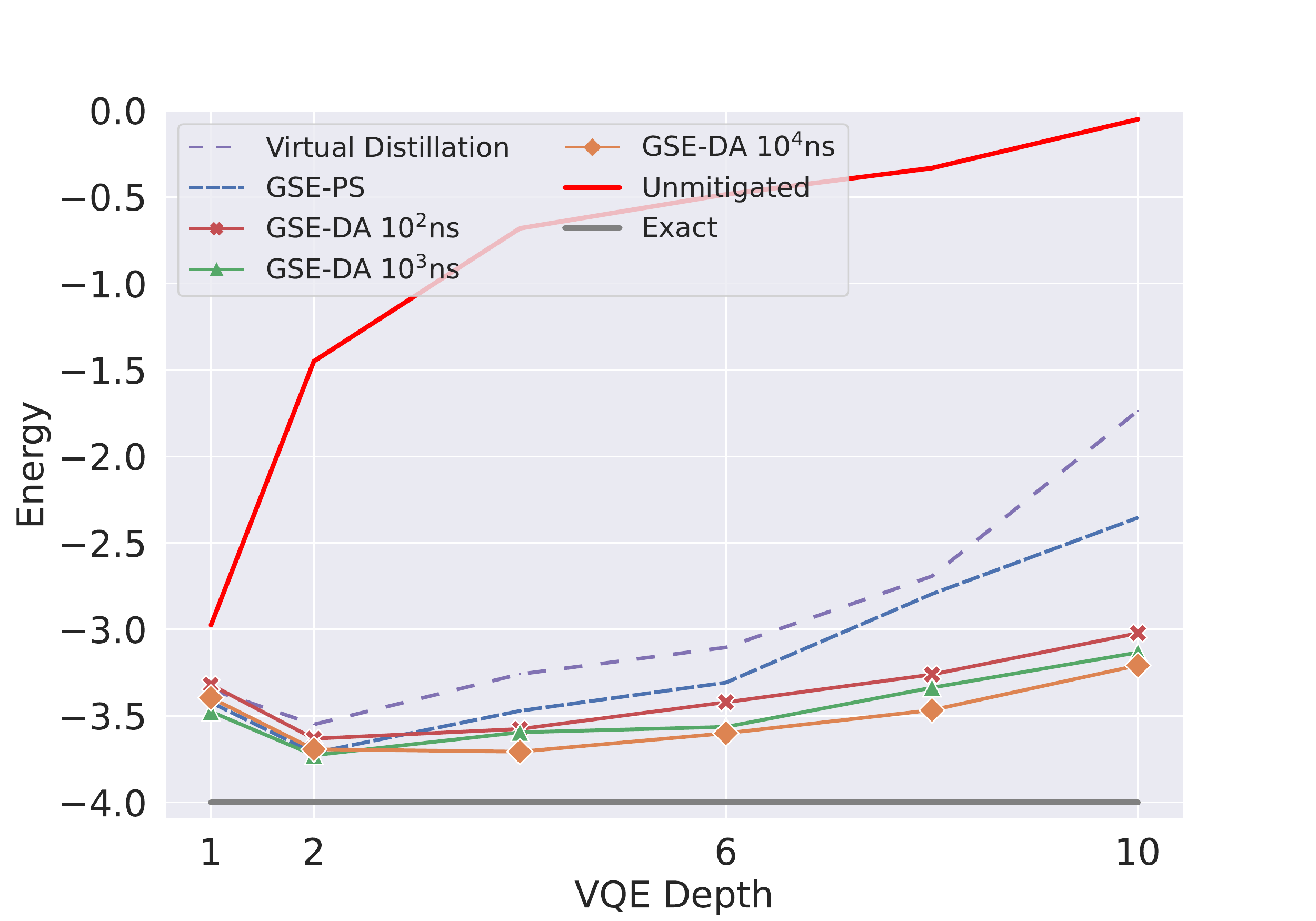}}
            \subcaption[]{Decoherence Amplification}
        \end{subfigure}
        \begin{subfigure}{.32\hsize}
            \resizebox{\hsize}{!}{\includegraphics{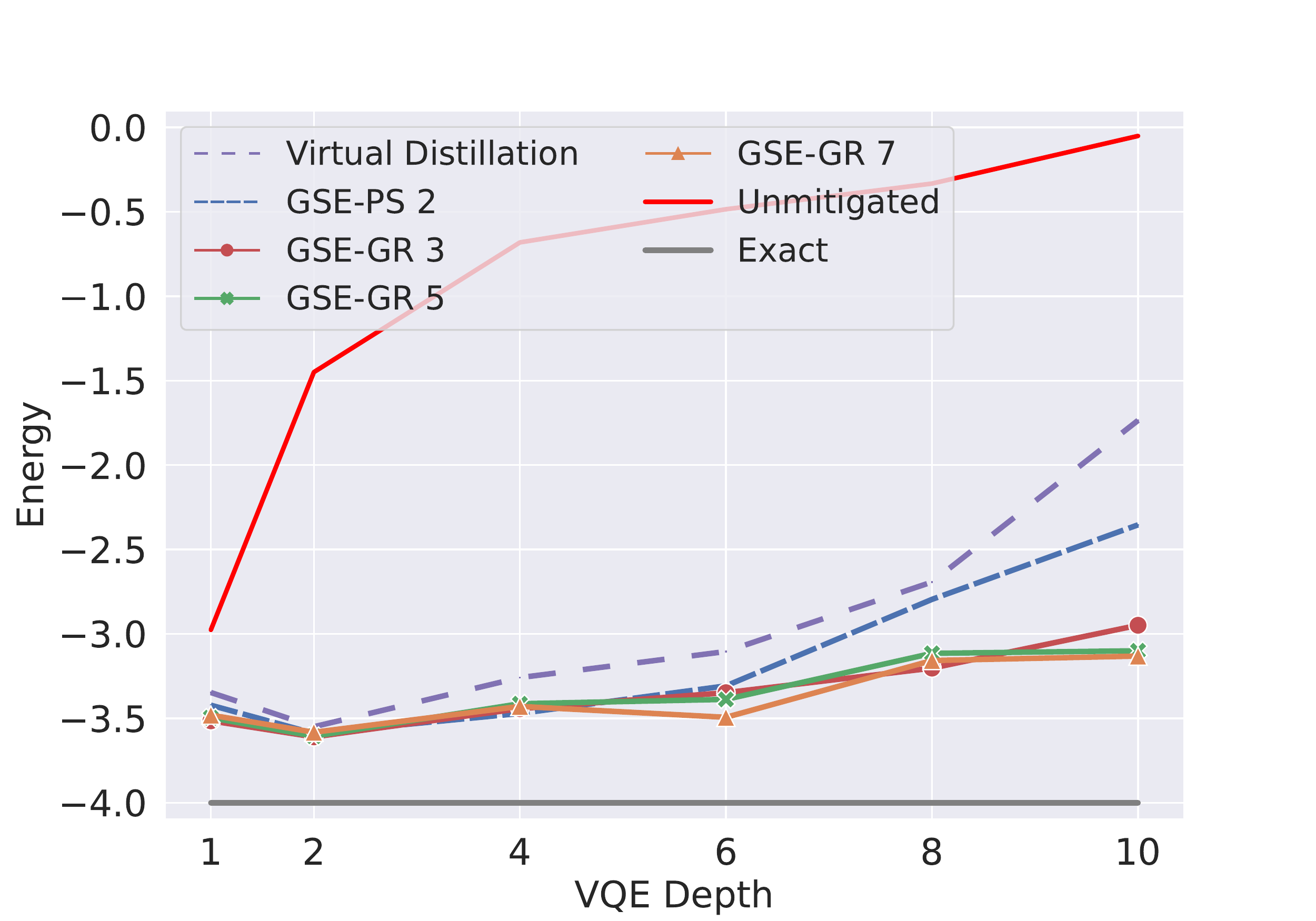}}
            \subcaption[]{Gate Repetition}
        \end{subfigure}
        \begin{subfigure}{.32\hsize}
            \resizebox{\hsize}{!}{\includegraphics{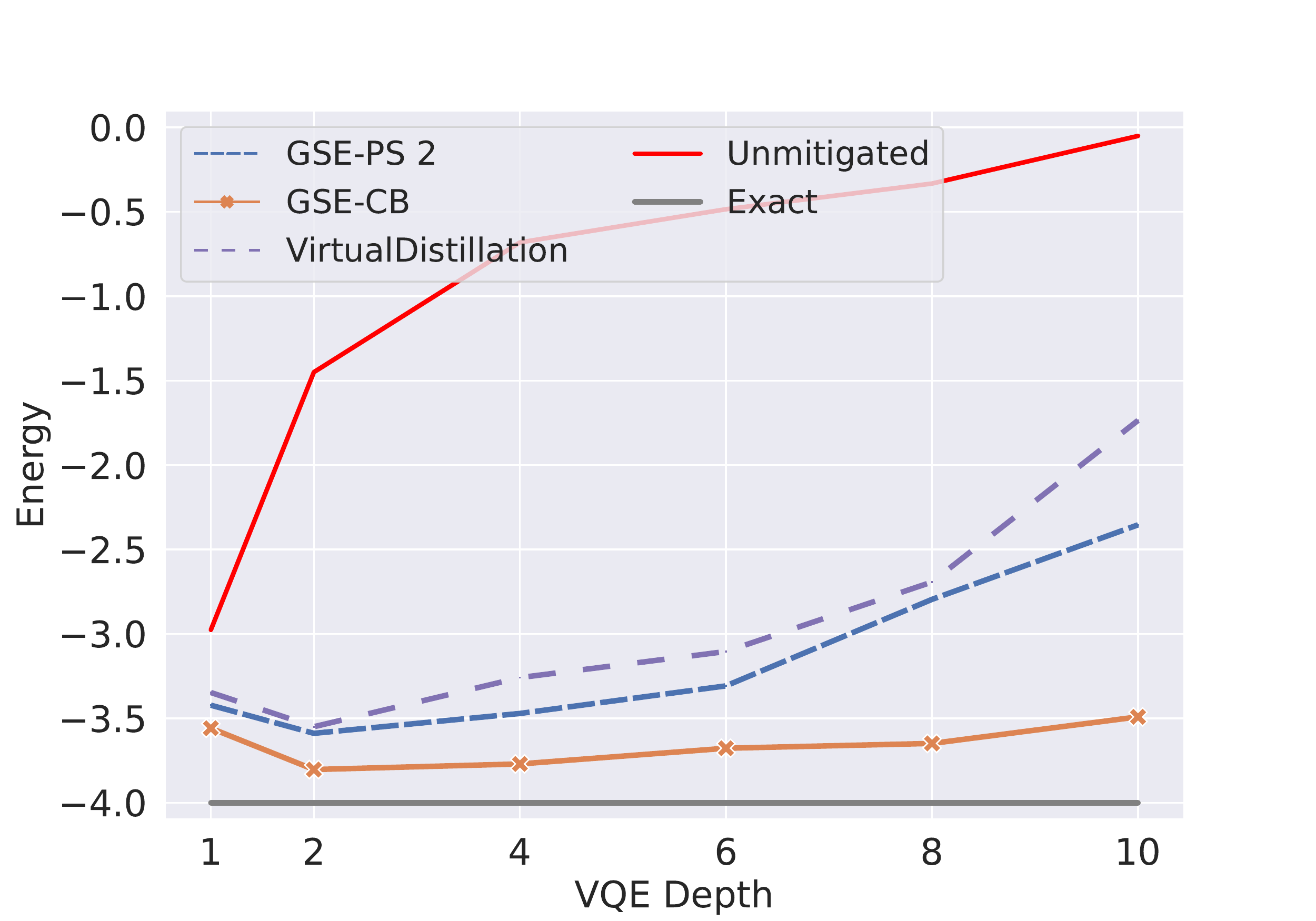}}
            \subcaption[]{Crosstalk Boost}
        \end{subfigure}
        \caption{
            Numerical demonstration of the error mitigation methods under various depths of the variational ansatz. Here we display the performance of GSE method using the fault subspace with the errors boosted by
            (a) decoherence amplification, (b) gate repetition, and (c) crosstalk boost.
            The black and red lines denote the exact ground state energy and the noisy value from the raw noisy circuit, respectively. For comparison, we also show the result from the VD method (purple dotted) and GSE with power subspace using $M=2$ copies (blue dashed).
            For the decoherence amplification technique, we compare the results with buffer time of $10^2, 10^3, 10^4$ ns and find that $10^4$ ns gives the best result here.
            Meanwhile, we did not find a significant difference between repetition counts for the gate repetition technique.
        }
    \label{fig:full_3_aer_simulator(ibm_kawasaki)_stretchfactor}
\end{figure*}

Figure~\ref{fig:full_3_aer_simulator(ibm_kawasaki)_stretchfactor} shows the expectation values of energy under various QEM methods. 
We find that all hardware-oriented GSE methods performs better than the vanilla VD method, while the separation of the performance turns out to be more significant in circuits with larger depth.
We can understand this from the ``effective dimension" of the fault subspace; the amplified decoherence gradually dominates compared to the other error sources that come from the circuit operation.
We find there is a certain sweet spot in the stretch factor of noise amplification. 
For instance, in Fig.~\ref{fig:full_3_aer_simulator(ibm_kawasaki)_stretchfactor}(a) we find that the noisy quantum state copy with additional decoherence time of $10^4$\si{\nano \second} shows the best performance among GSE-DA techniques.

Given that the current noisy simulation is based on the error profile on quantum devices \textit{ibm\_kawasaki}, it is natural to expect that the hardware-oriented GSE methods will perform quite robustly even under \textcolor{black}{demonstration} on quantum hardware.
\subsection{Demonstration on the quantum device}
\label{subsec:experiments}

\begin{figure*}[t]
    \centering
        \begin{subfigure}{.45\hsize}
            \resizebox{\hsize}{!}{\includegraphics{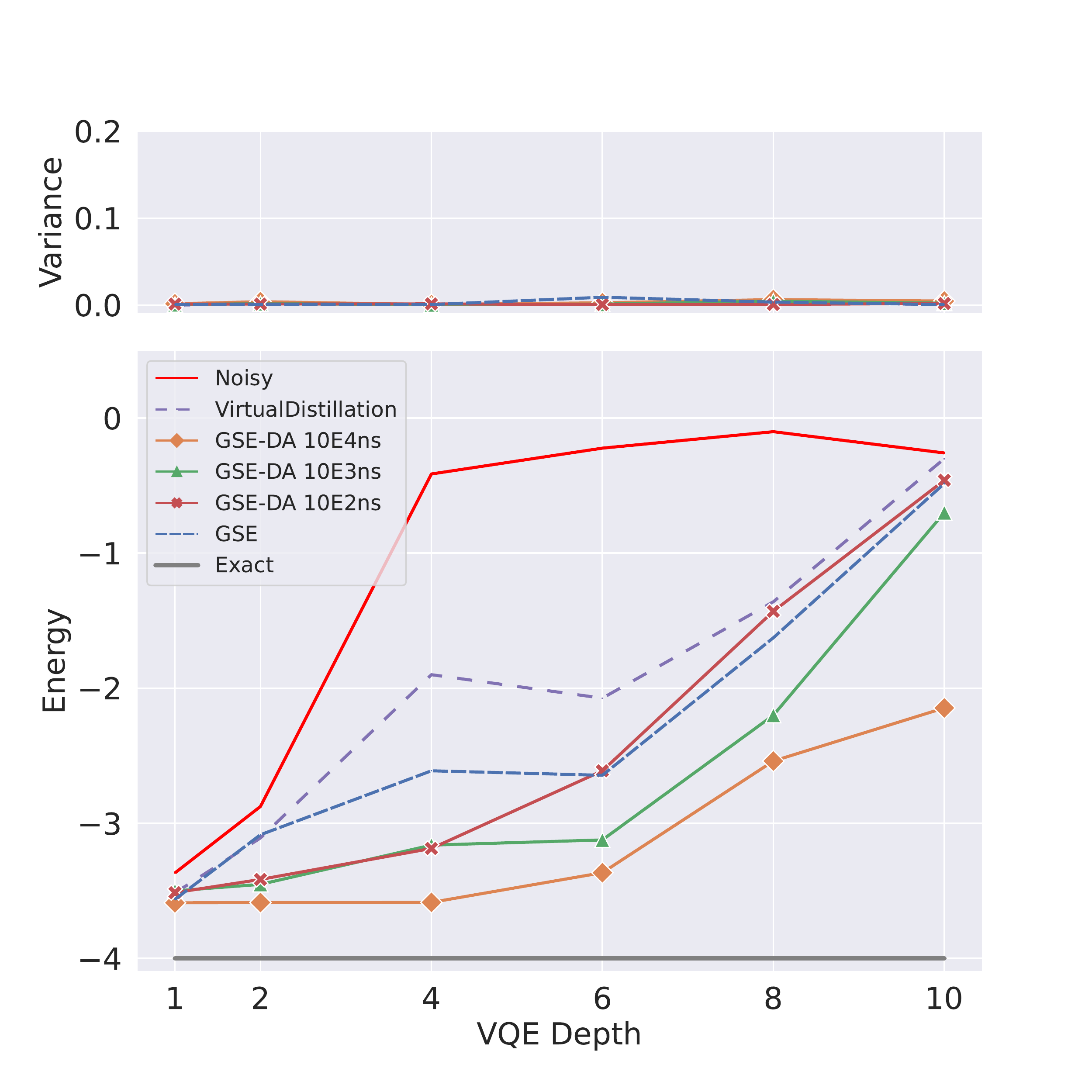}}
            \subcaption[]{3-qubits size Hamiltonian}
        \end{subfigure}
        \begin{subfigure}{.45\hsize}
            \resizebox{\hsize}{!}{\includegraphics{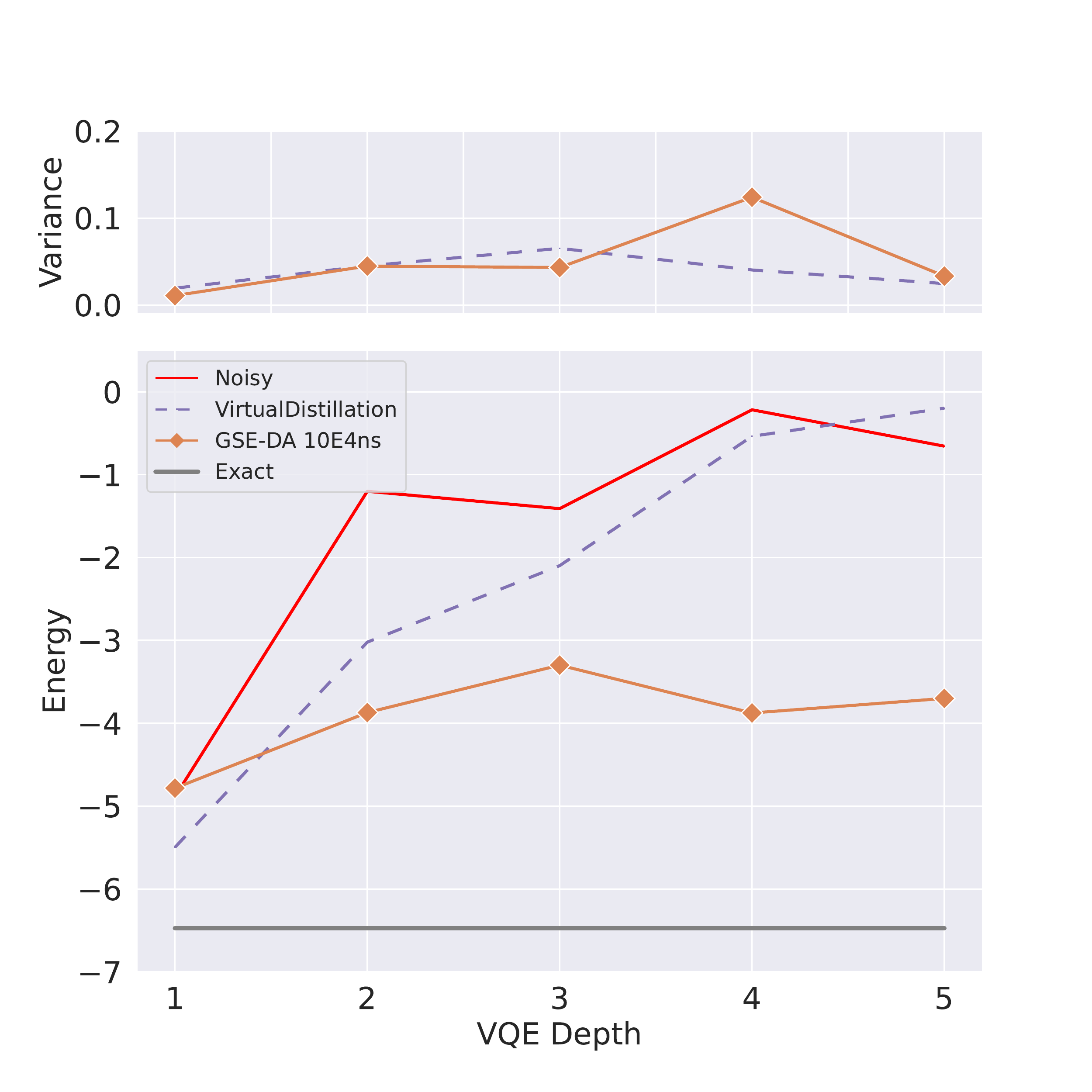}}
            \subcaption[]{5-qubits size Hamiltonian}
        \end{subfigure}
        \caption{
            Evaluation of scalability on current Processor. 
            We conduct the same \textcolor{black}{demonstrations} with larger qubit sizes. Due to utilizing 2 copies of quantum state $\rho$ in subspace, our technique works for in total of 10-qubit size circuits at maximum.
            In the 3-qubit problem, we vary the duration time to introduce decoherence of quantum state from $10^2$\si{\nano \second} to $10^4$\si{\nano \second}. 
            The ideal expectation values are $-4$ and $-6.47$ with respect to the 3-qubit and 5-qubit size problems. 
        }
    \label{fig:result_on_realdevice}
\end{figure*}

Next, we present the results of demonstration that are performed on
 $27$-qubit quantum processor, \textit{ibm\_kawasaki}, provided by IBM Quantum (See Appendix~\ref{app:topology}  for qubit topology).
In particular, we verify the validity of GSE-DA method by simulating the ground state of the 1d TFI model with $h=1$ as well as in the numerical simulation.

Figure~\ref{fig:result_on_realdevice} shows the result on the \textit{ibm\_kawasaki} device to compute expectation values for 3-qubit and 5-qubit sized problems.
We find that in both cases we observe a significant separation of accuracy between GSE-DA method and the raw output.
Reflecting the fact that GSE-DA in the current setup uses two copies of the noisy quantum state, the data follows a similar trend as in the VD method which simply purifies the noisy state without correcting any coherent errors.
Such a difference yields better performance of GSE-DA method, which is evident in deeper circuits where errors in gates are expected to introduce drift in the dominant vector (or the coherent mismatch~\cite{koczor2021dominant}) of the density matrix.

We remark that we have designed the variational circuit (See Appendix~\ref{app:ansatz}) to involve an excessive number of entangling gates so that the effect of the noise is more severe;
the number of CX gates are $O(N^2 d)$ for an $N$-qubit system.
This is reflected in the fast drop of the accuracy of the noisy raw energy and the VD method in $N=5$ qubit simulation shown in Fig.~\ref{fig:result_on_realdevice}(b). Meanwhile, we observe that GSE-DA method in the noisy regime yields significantly better performance due to the construction of the variational subspace in the postprocessing. 

\section{Conclusion}
\label{sec:conclusion}
In this work, we proposed novel techniques to leverage hardware imperfections to simulate eigenstates of quantum many-body systems by extending the framework of the fault subspace for the generalized quantum subspace expansion method. 
\black{The validity of the proposed method is verified via both numerical simulations and demonstrations on a quantum device. In particular, the flavor of the fault subspace using the decoherence amplification, which we have denoted as GSE-DA, has shown stability over the vanilla purification-based error mitigation method.}

\black{
We envision two main directions as future problems.
First, it may be highly beneficial to seek how to lower the overhead in the entangled measurement associated with the derangement operation.
At the hardware level, we point out that it is important to utilize a quantum processor with higher qubit connectivity so that derangement operation can be performed with smaller overhead. 
At the algorithmic level, we may even consider reducing the number of entangled measurements; in the case of a quantum state with relatively weak quantum entanglement, we expect that the simulation is well-approximated even if we perform the derangement operation  only locally on sites where observable have support on. 
}

\black{
Second, it is interesting to investigate how to systematically find the efficient construction of the erroneous subspace.
While we have heuristically enumerated various sources of errors that may contribute to expanding the variational subspace for GSE method, we emphasize that it is a central problem among all subspace-based methods to establish a guiding principle to choose the base.
Hence, such a further investigation may benefit not only error mitigation but also vast fields such as computational physics and quantum chemistry.
}

{\it Acknowledgements.---}
S.E wishes to thank PRESTO, JST, Grant No.\,JPMJPR2114; CREST, JST, Grant No.\,JPMJCR1771; MEXT Q-LEAP Grant No.\,JPMXS0120319794 and JPMXS0118068682, JST Moonshot R\&D, Grant No.\,JPMJMS2061. 
T.S  wish to thank JPMXS0118067285 Q-LEAP, JPMXS0120319794 Q-LEAP and the MEXT KAKENHI Grant Number 22K19781.
rdv wish to thank JPMXS0118067285 Q-LEAP and JPMXS0120319794 Q-LEAP.
N.Y. wishes to thank JST PRESTO No. JPMJPR2119 and the support from IBM Quantum.
This work was supported by JST Grant Number JPMJPF2221.
We acknowledge the use of IBM Quantum services for this work. The views expressed are those of the authors, and do not reflect the official policy or position of IBM or the IBM Quantum team.

\appendix
\section{Noise channels for numerical simulation}\label{app:noise}

It is informative to explicitly write down the effect of noise channels considered in the numerical simulations.
In particular, here we assume that local quantum gates are all exposed to local noise. Under 
$T_1$ and $T_2$ relaxations, for instance, we introduce noise channel as $\mathcal{E} = \otimes_{n=1}^N \mathcal{E}_{n}^{(T_1/T_2)}$ where $\mathcal{E}_{n}$ denotes the local error channel acting on the $n$-th qubit.
The expression of two noises $\mathcal{E}^{(T_1)}_n$ and $\mathcal{E}^{(T_2)}_n$ are given as
\begin{eqnarray}
        \mathcal{E}_n^{(T_1)}\left(\cdot\right) &=& K_{n,0} (\cdot) K_{n,0}^\dagger + K_{n,1} (\cdot) K_{n,1}^\dagger,\\
        \mathcal{E}_n^{(T_2)}\left(\cdot\right) &=& 
        K_{n,2}(\cdot)K_{n,2} + K_{n,3}(\cdot)K_{n,3}.
\end{eqnarray}
With the Pauli operators on $n$-th site denoted as $I_n, X_n, Y_n,$ and $Z_n$, the Kraus operators can be expressed as
\begin{eqnarray}
K_{n,0}&=& \frac{1+\sqrt{1 - e^{-t/T_1}}}{2}I_n + \frac{1-\sqrt{1-e^{-t/T_1}}}{2}Z_n, \\
K_{n,1} &=& e^{-t/2T_1}\frac{X_n + i Y_n}{2}, \\
    K_{n,2} &=& e^{-t/2T_2}I_n, \\ 
    K_{n,3} &=& (1-e^{-t/2T_2}) Z_n,
\end{eqnarray}
where $t$ is the execution time of the quantum gate, $T_1$ and $T_2$ are the relaxation time.
For instance, the duration for performing gate operation of \textit{ibm\_kawasaki} on the date we conducted \textcolor{black}{demonstration}s was $366.2$ns for CX gate on average.
The representation of the Kraus operators for $T_1$ effect can be given alternatively using the tensor-product-wise description as
\begin{eqnarray}
        K_{n,0} &=& I\otimes \cdots \otimes \left(\begin{array}{ll}
            1 & 0 \\
            0 & \sqrt{1-e^{-t / T_{1}}}
        \end{array}\right)\otimes \cdots \otimes I, \\
        K_{n,1} &=& I\otimes \cdots \otimes  \left(\begin{array}{ll}
        0 & \sqrt{e^{-t /  T_{1}}} \\
        0 & 0
        \end{array}\right)\otimes \cdots \otimes I.
\end{eqnarray}
Note that values $T_1$ and $T_2$ can be determined by, e.g., measuring the lifetime of the excited state and Hahn echo experiment, respectively.

\section{Qubit topology of the hardware} \label{app:topology}
\begin{figure}[t]
    \centering
    \resizebox{0.9\hsize}{!}{\includegraphics[]{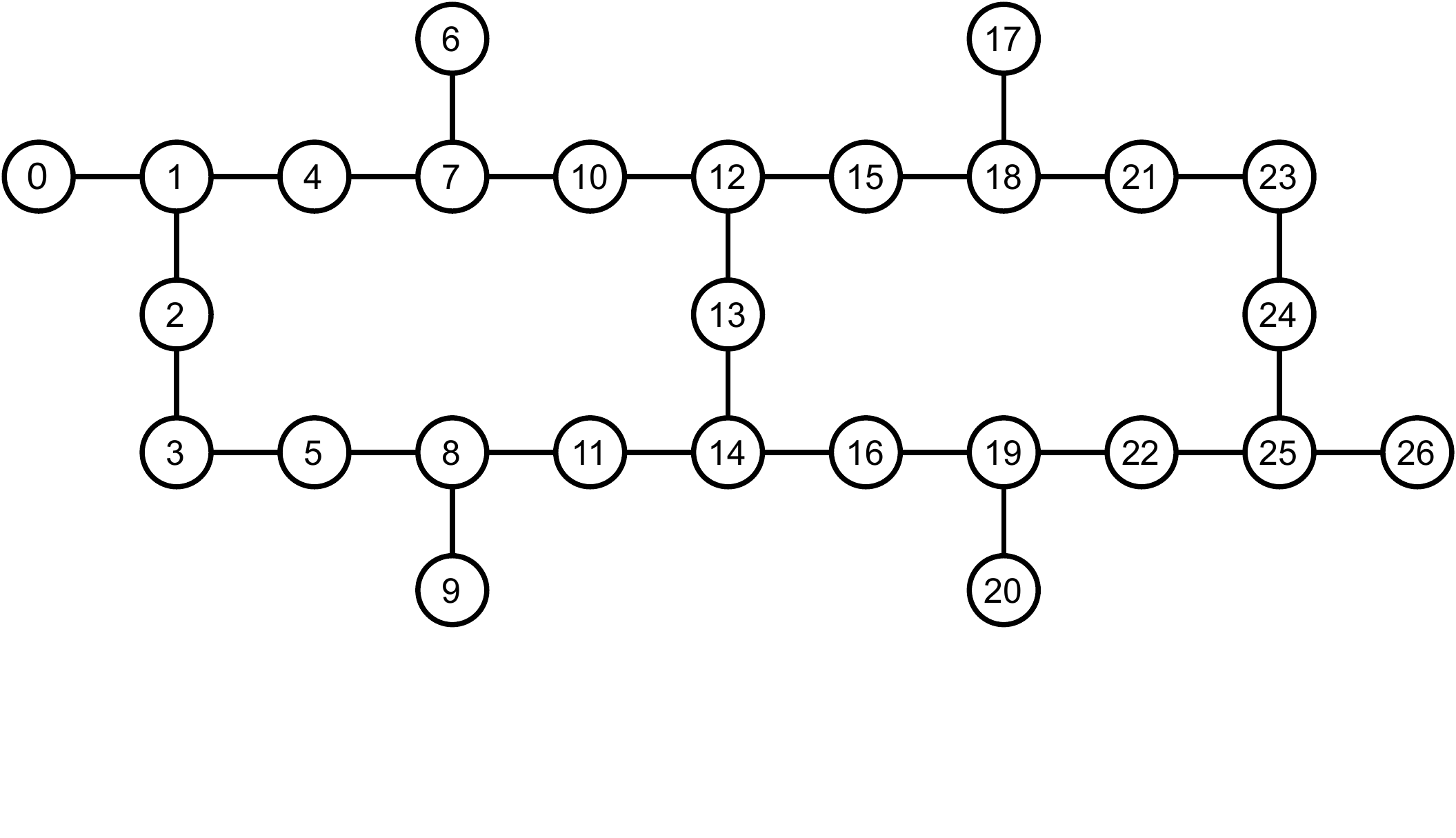}}
    \caption{
        Qubit topology of \textit{ibm\_kawasaki} which exhibit quantum volume of 128~\cite{ibm_quantum_experience}.
        Each qubit has a label for convenience.
    }
    \label{fig:ibm_kawasaki}
\end{figure}

Here we show in Fig.~\ref{fig:ibm_kawasaki} the qubit connectivity of {\it ibm\_kawasaki}, which is the current common quantum device available via the cloud service provided by IBMQ.
Authorized readers may access further details on the device such as qubit frequency or gate fidelity via the website of IBM Quantum~\cite{ibm_quantum_experience}.

\section{Variational ansatz for ground state simulation} \label{app:ansatz}
\begin{figure}[t]
    \begin{center}
    \centering
    \resizebox{0.95\hsize}{!}{\includegraphics[]{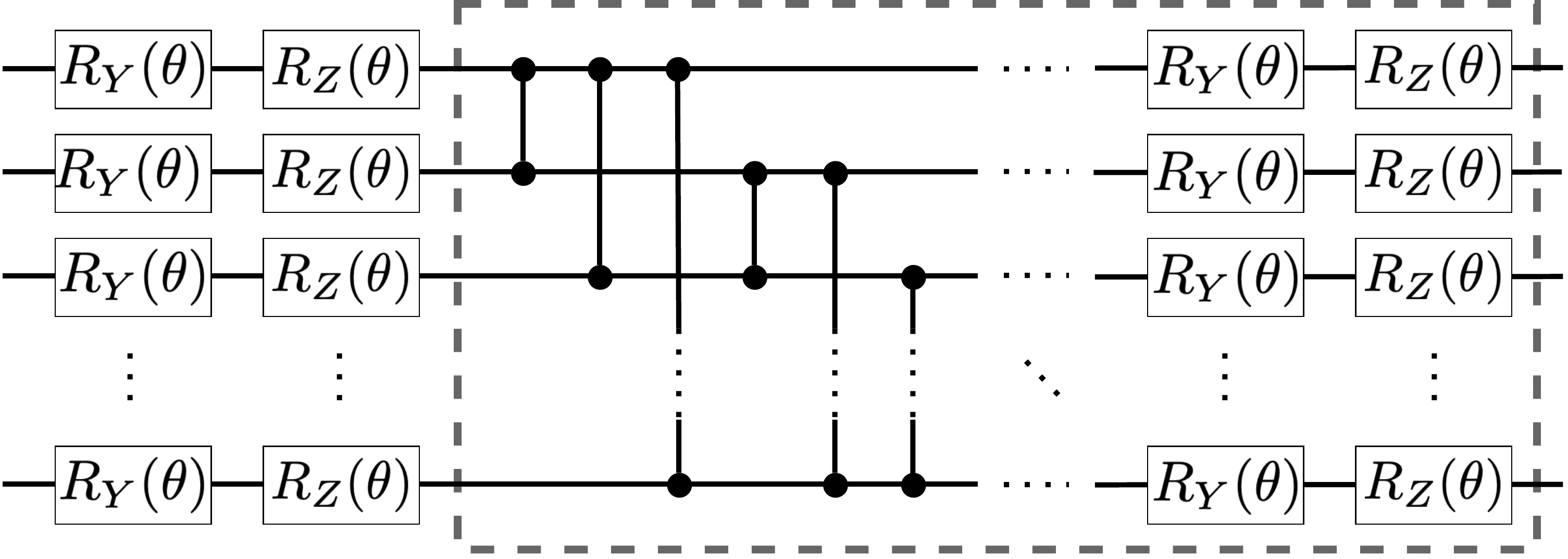}}
    \caption{
        Structure of the variational ansatz considered in our demonstration.
        We employ a hardware-efficient fully-entangled structure, in which CZ gates are applied to  all the two-qubit pairs of the state.
        The gates inside the dashed line are repeated $d$ times where $d$ is referred to as the depth. 
        In the \textcolor{black}{demonstrations}, we use 3-qubit size variational circuits, depth from 1 to 10 both in simulation and quantum device \textcolor{black}{demonstration}s, and 5, 7, and 9 qubits size circuits with depths 1 to 5.  
    }
    \label{fig:variational_circuit}
    \end{center}
\end{figure}

Figure~\ref{fig:variational_circuit} shows the circuit structure of the variational ansatz used for the numerical and experimental demonstration of the hardware-oriented GSE method for the ground state calculation.
The all-to-all connected structure allows the ansatz to capture larger quantum correlation via the number of variational parameters scaling as $2N(d+1)$ where $N$ is the number of qubits and $d$ is the repetition number (or ``depth") of the block denoted by the dashed line.

\section{Implementation of derangement operator with linear connectivity}

\label{sec:implementation}
In order to perform purification-based error mitigation on hardware that does not support all-to-all connectivity, we must reorder qubits in order to perform the entangled measurement.
Such reordering introduces additional overhead on top of the basis transformation required for the entangled measurement itself.
In this section, we discuss how to implement the qubit ordering with the minimum requirement on the hardware; the linear connectivity.

In particular, we focus on the number of CX gates since two-qubit gates are a major resource of errors in various quantum hardware.
For instance, let us assume that the total number of CX gates $G_{tot}$ in the  quantum circuit under all-to-all connectivity can be provided as
\begin{eqnarray}
G_{tot} = G_{\rm VQE} + G_{derange}
\end{eqnarray}
where $G_{\rm VQE}$ and $G_{derange}$ are the number of CX gates required to implement the variational ansatz for the VQE algorithm and basis transformation $B_\sigma^{\otimes N}$ to perform the entangled measurement.
However, in reality, there is an additional overhead due to the restricted qubit topology. In other words, we must compile, or solve the routing task, so that the sequence of gates acting on the given restricted topology gives identical computation with the original precompiled quantum circuit.

\begin{figure}[ht]
    \centering
        \resizebox{0.95\hsize}{!}{\includegraphics{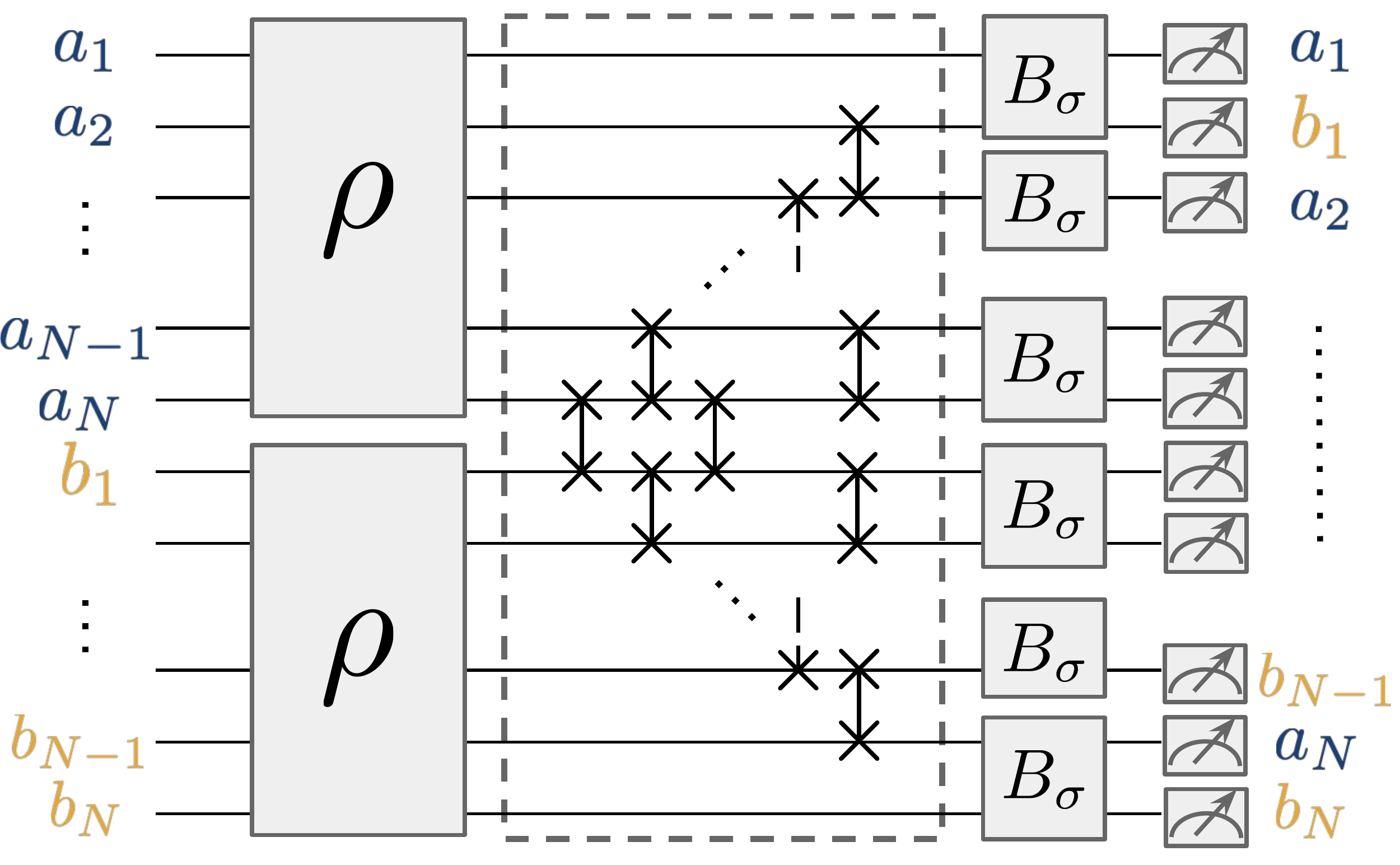}}
    \caption{
        Graphical description of the alternating SWAP technique to rearrange the qubit ordering so that the entangled measurement can be executed on hardware with linear connectivity.
    }
    \label{fig:alternating_swap}
\end{figure}

Here, we introduce a mapping/routing technique that we call alternating SWAP. 
While we here focus on the description for the case of $M=2$ copies, it can be straightforwardly extended for multiple $M$ as well.
Let $a_k$ and $b_k$ denote the $k$-th qubits of the first and second copies of quantum states, respectively.
The alternating SWAP technique is designed so that the sequence of qubits $\{a_1, ..., a_N, b_1, ..., b_N\}$ are rearranged so that corresponding sites are near to each other as $\{a_1, b_1, ..., a_N, b_N\}$(See Fig.~\ref{fig:alternating_swap}).
To be concrete, the procedure of alternating SWAP is described as follows:
\begin{enumerate}
    \item[Step 1.] We prepare copies of noisy state $\rho_{m}$ and map next to each other.
     \item[Step 2.] Execute a hierarchical SWAP operation to rearrange the qubit ordering so that qubits denoting the identical label are located next to each other.
    \item[Step 3.] Perform the entangled measurement by operating the basis transformation unitary $B_\sigma$ followed by the ordinary single-site measurement.
\end{enumerate}
Summed up together with the original quantum circuit, the number of two-qubit CX gates in the compiled quantum circuit is now given as
\begin{eqnarray}
G_{tot} = G_{\rm VQE} + \frac{N(N+1)}{2}G_{\rm swap} + G_{derange},
\end{eqnarray}
where $G_{\rm swap}=3$ in the current work.

\subsection{Cost of alternating SWAP technique}
\begin{figure}[t]
    \begin{center}
    \resizebox{0.95\hsize}{!}{\includegraphics{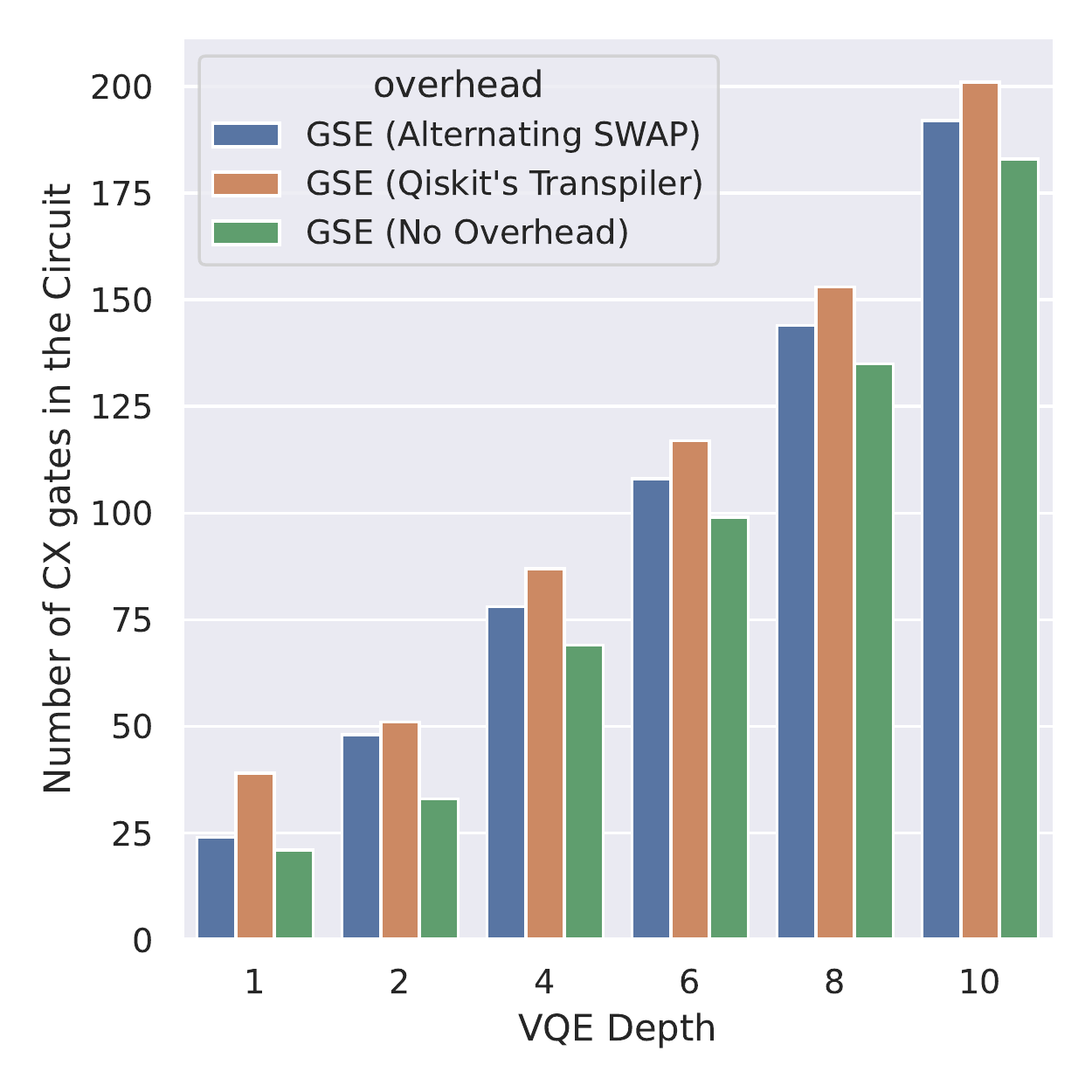}}
        \caption{
            CX gate count in 6-qubit size Quantum Circuit (2 copies of 3-qubit size quantum states).
            Here, 
            \textit{GSE (No Overhead)} denotes the CX gate count assuming all-to-all connectivity of the quantum circuit to implement the fault subspace.
            \textbf{GSE (Alternating SWAP)} and \textbf{GSE (Qiskit's Transpiler)} show the CX gate count by further taking into account the limitation of the connectivity in heavy-hexagon structure.
        }
    \label{fig:overheads_of_circuit_alternateSWAP}
    \end{center}
\end{figure}

Here we evaluate the overhead of introducing the alternating SWAP technique.
Figure~\ref{fig:overheads_of_circuit_alternateSWAP} shows the number of CX gates required to perform GSE method with $M=2$ copies for  $N=3$ qubit system.
Compared with the overhead obtained by performing transpilation solely using the function of Qiskit~\cite{Qiskit}, we observe that our mapping technique always provides smaller overhead no matter how deep the variational circuit is. 
This is somewhat expected since the alternating SWAP technique is involved only after the variational gates are operated.

\begin{figure}[ht]
    \begin{center}
    \resizebox{0.95\hsize}{!}{\includegraphics{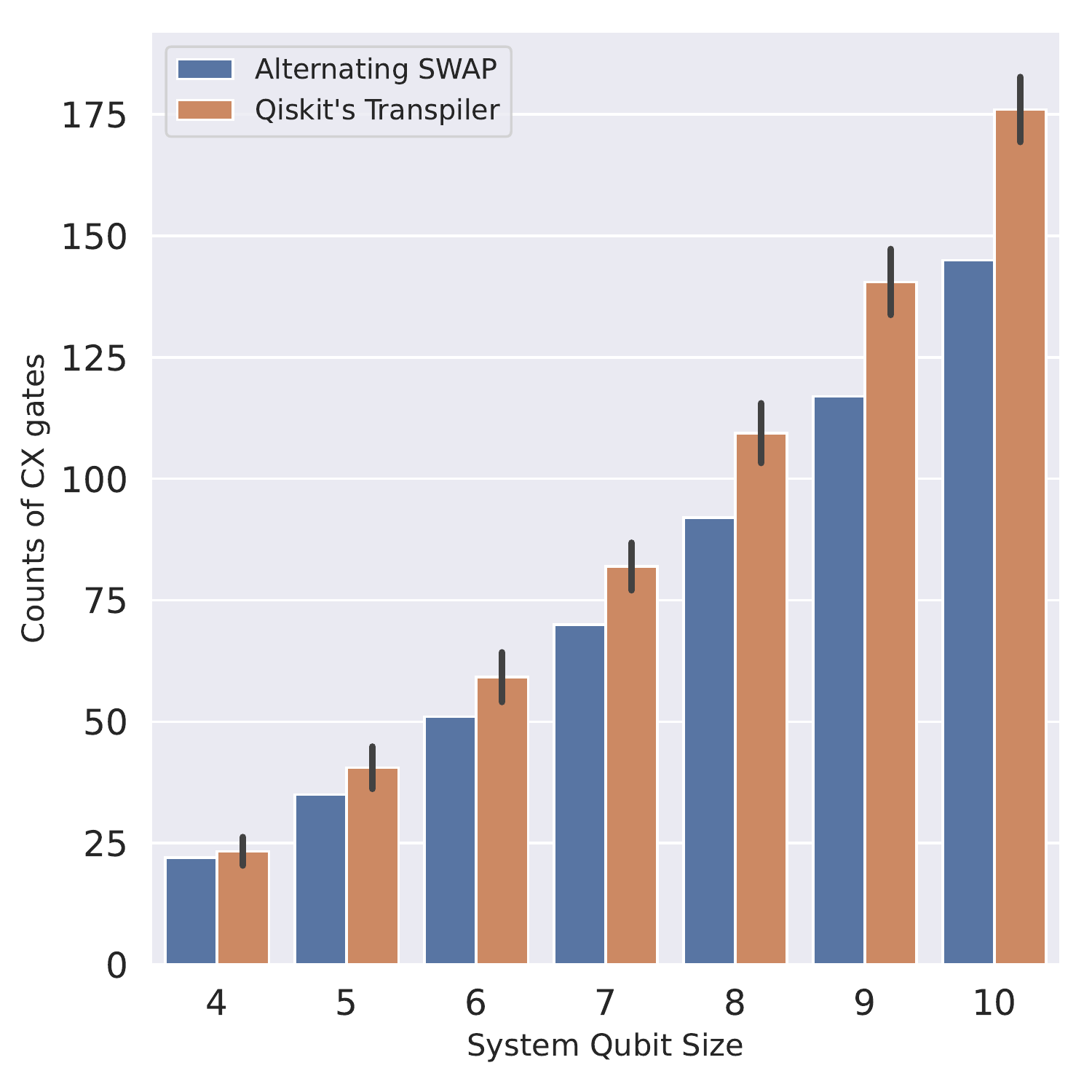}}
        \caption{
            CX gate count overhead for various system sizes.
            The alternating SWAP technique always outperforms the transpiler provided by Qiskit.
        }
    \label{fig:gate_counts}
    \end{center}
\end{figure}

Fig.~\ref{fig:gate_counts} shows the overhead under various system sizes.
We show the gate counts up to 10-qubit system size, i.e. it occupies 20-qubit of the quantum processor.
We find that the alternating SWAP technique achieves routing with smaller overhead at any system size considered in numerical simulation. We expect that it is optimal in terms of CX gate count for arbitrary size.

\subsection{Effect of qubit ordering overhead on simulation accuracy}
\begin{figure}[ht]
    \begin{center}
    \resizebox{0.95\hsize}{!}{\includegraphics{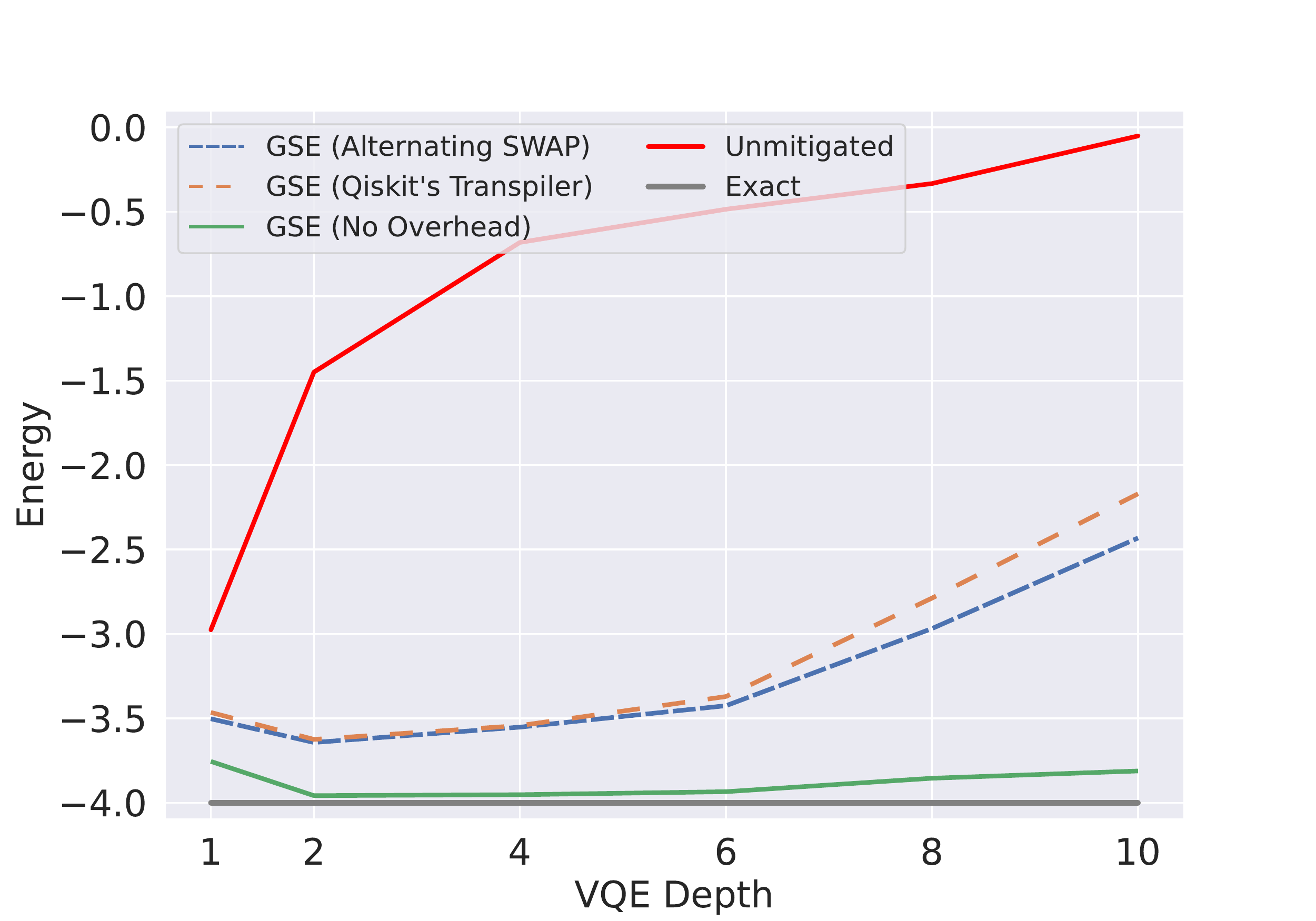}}
        \caption{
            Comparison of simulation accuracy with and without overhead regarding qubit ordering overhead.
            Here, we simulate GSE method with power subspace using $M=2$ copies on $N=3$ qubits.
            A green line denotes the case without any ordering overhead (i.e. all-to-all connectivity), whereas 
            orange and blue lines represent the circuit compiled by Qiskit's transpiler and alternating SWAP technique, respectively. 
            The graph shows the result with the alternating SWAP technique provides better accuracy for deeper circuits.
        }
    \label{fig:performance_of_alternating_SWAP}
    \end{center}
\end{figure}

Here we discuss the effect of overhead by qubit ordering on the accuracy of the ground state simulation.
We compiled the QEM circuit for VD and GSE-DA methods to resolve constraints of the quantum device \textit{ibm\_kawasaki} (such as physical two-qubit connection, the error rate of qubits, gate time, etc.) by adding gate operations.
Fig.~\ref{fig:performance_of_alternating_SWAP} shows the comparison of the performance of GSE method with power subspace using $M=2$ copies with/without qubit ordering overhead.
The case of no-overhead mitigates the noise nearly perfectly and that is the almost identical result as the original paper demonstrated.
Compared to that, both results of compiled circuits are much worse.
Our proposal for the alternating SWAP technique suppresses the noise effect and shows a better solution than Qiskit compilation in the range of deeper circuits.
We envision that such a deterioration of the QEM simulation can be alleviated by using the dual form of VD method, for instance~\cite{huo2021dual}.

\bibliography{bib_GSE}
\end{document}